\documentclass[twocolumn,a4paper,english]{revtex4}
\usepackage{ae,color,graphicx}
\usepackage[latin1]{inputenc}
\usepackage[T1]{fontenc}
\usepackage[english]{babel}

\begin{document}

\title{Event-by-Event Analysis of Baryon-Strangeness Correlations:\\ 
Pinning Down the Critical Temperature and Volume of QGP Formation}
\author{Stephane Haussler$^{1}$, Horst Stöcker$^{1,2}$, and Marcus Bleicher$^{2}$}
\affiliation{1) Frankfurt Institute for Advanced Studies (FIAS),
Johann Wolfgang Goethe Universität, Max-von-Laue-Str. 1,
60438 Frankfurt am Main, Germany\\
2) Institut für Theoretische Physik, Johann Wolfgang Goethe
Universität, Max-von-Laue-Str. 1, 60438 Frankfurt am Main, Germany }

\begin{abstract}
The recently proposed baryon-strangeness correlation ($C_{BS}$) is studied
with a string-hadronic transport model (UrQMD) for various energies from $E_{\rm lab}=4$~AGeV to
$\sqrt s=200$~AGeV. It is shown that rescattering among secondaries
can not mimic the predicted correlation pattern expected for a Quark-Gluon-Plasma.
However, we find a strong increase of the $C_{BS}$ correlation function
with decreasing collision energy both for pp and Au+Au/Pb+Pb reactions.
For Au+Au reactions at the top RHIC energy ($\sqrt s=200$~AGeV),
the $C_{BS}$ correlation is constant for all centralities and compatible
with the pp result. With increasing width of the rapidity window, $C_{BS}$ follows
roughly the shape of the baryon rapidity distribution.
We suggest to study the energy and centrality dependence of $C_{BS}$ which allow to 
gain information on the onset of the deconfinement transition in temperature and volume.
\end{abstract}
\maketitle

Several observables \cite{Bass:1998vz} have been proposed throughout the last decades
to study the characteristics of the highly excited matter
created in heavy ions collisions, where a Quark-Gluon Plasma (QGP) is believed to be created.
Among these observables, that give the opportunity to probe whether
or not the system went through a phase of deconfined quarks and gluons,
the ones related to fluctuations and correlations
seem to be the most prospective. Fluctuation probes might be more adequate for the
exploration of heavy ion reactions, because the distributions of energy density
or initial temperature, isospin and particle density have strong fluctuations
from event to event \cite{Stodolsky:1995ds,Shuryak:1997yj,Bleicher:1998wd}.
On the theoretical side event-by-event fluctuations where suggested to study
\begin{itemize}
\item
kinetic and chemical equilibration in nuclear collisions \cite{Gazdzicki:1992ri,Mrowczynski:1997kz,Bleicher:1998wu,Belkacem:1999ui,Mrowczynski:1999sf,Capella:1999uc,Mrowczynski:1999un,Mrowczynski:2004cg,Sa:2001ma,Mekjian:2004qf},
\item
the onset of the deconfinement phase \cite{Asakawa:2000wh,Muller:2001wj,Bleicher:2000ek,Koch:2001zn,Jeon:2003gk,Shi:2005rc,Jeon:2005kj}
\item
the location of the tri-critical end-point of the QCD phase transition \cite{Stephanov:1998dy,Stephanov:1999zu,Hatta:2003wn} or 
\item
the formation of exotic states, like DCCs \cite{Bleicher:2000tr}.
\end{itemize}
On the experimental side, progress has been made by many experiments
to extract momentum and particle number ratio fluctuations from
heavy ion reaction:  E-by-E fluctuations are actively studied in the SPS energy regime (starting from 20~AGeV on)
by the NA49 group \cite{Appelshauser:1999ft,Reid:1999it,Afanasev:2000fu,Anticic:2003fd,Roland:2004pu,Alt:2004ir,Alt:2004gx,Rybczynski:2004yw,Roland:2005pr} and the CERES  collaboration \cite{Adamova:2003pz,Sako:2004pw,Appelshauser:2004xj,Appelshauser:2004ms}.
At RHIC energies the PHENIX \cite{Adcox:2002mm,Adcox:2002pa,Adler:2003xq} and 
STAR \cite{Adams:2003st,Adams:2003uw,Adams:2004kr} experiments 
are addressing the field of single event physics. 

Recently a novel event-by-event observable has been introduced by Koch et al. \cite{Koch:2005vg}, 
the baryon-strangeness correlation coefficient $C_{BS}$. 
This correlation is proposed as a tool to specify the nature (ideal QGP or strongly coupled QGP or hadronic matter) 
of the highly compressed and heated matter created in heavy ions collisions.
The idea is that depending on the phase the system is in, the relation between 
baryon number and strangeness will be different:
On the one hand, if one considers an ideal plasma of quarks and gluons,
strangeness will be carried by freely moving strange and anti-strange quarks,
carrying  baryon number in strict proportions.
This leads to a strong correlation between the baryon number and strangeness.
On the other hand, if the degrees of freedom are of hadronic nature, this correlation 
is different, because it is possible to carry strangeness without baryon number, e.g. in mesons or QGP bound states.

To quantify to which degree strangeness and baryon number are correlated,
the following correlation coefficient has been proposed \cite{Koch:2005vg}:

\begin{equation}\label{eq:definition}
C_{BS}=-3 \frac{\langle BS\rangle-\langle B\rangle\langle S\rangle}{\langle S^{2}\rangle-\langle S\rangle^{2}}\quad,
\end{equation}
where $B$ is the baryon charge and $S$ is the strangeness.
If a QGP is created, the expected value of $C_{BS}$ will be unity as expected from lattice QCD, 
compatible with the ideal weakly coupled QGP.
In the case of a hadron gas, where the correlation is non trivial,
this quantity has been evaluated in  \cite{Koch:2005vg} to be $C_{BS}=0.66$.

In this paper, we study the correlation coefficient $C_{BS}$
with the Ultra-relativistic Quantum Molecular Dynamics model (UrQMD v2.2).
The UrQMD is a non-equilibrium microscopic transport model that simulates the full space-time evolution of heavy ions collisions.
It is valid from a few tens of MeV to several TeV per nucleons in the laboratory frame.
It describes the rescattering of incoming and produced particles,
the excitation and fragmentation of color strings
and the formation and decay of resonances.
This model has been used before to study event-by-event fluctuations rather successfully \cite{Bleicher:1998wd,Bleicher:1998wu,Bleicher:2000ek,Bleicher:2000tr,Jeon:2005kj}  and yields a 
reasonable description of inclusive particle distributions.
For a complete review of the model, the reader is referred to \cite{Bass:1998ca,Bleicher:1999xi}.

Since the UrQMD is based on hadrons and strings it provides an estimate of the $C_{BS}$ value in 
the case where no QGP is created, however taking into account  the rescattering and the non-equilibrium nature
of the heavy ion reactions.
$C_{BS}$ is evaluated from the  event-by-event fluctuation analyses following \cite{Koch:2005vg,kochpriv}:

\begin{equation}\label{eq:calcul}
C_{BS}=-3 \frac{ \frac{1}{N} \sum_{n} B^{(n)} S^{(n)} - (\frac{1}{N} \sum_{n} B^{(n)}) (\frac{1}{N} \sum_{n} S^{(n)}) }
    {\frac{1}{N} \sum_{n} (S^{(n)})^2 - (\frac{1}{N} \sum_{n} S^{(n)})^2}
\end{equation}

$B^{(n)}$ and $S^{(n)}$ stand for the baryon number and strangeness in a given event $n$.

\begin{figure}
\begin{center}
\includegraphics[width=0.5\textwidth]{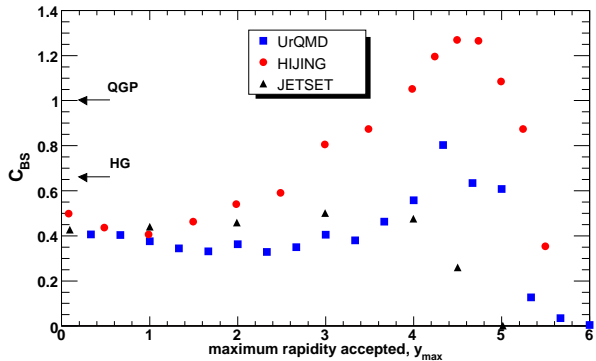}
\end{center}
\caption{Correlation coefficient for central Au+Au collisions at $\sqrt{s}=200$ shown as a function of the 
	maximum rapidity accepted.
	Circles are the calculation with HIJING \cite{Wang:1991ht}.
	Squares are the result of the UrQMD calculation and triangles of the JETSET for $e^+e^-$ at $\sqrt{s}=200$.
	The arrows are the values of a quark gluon plasma and of an hadron gas at a temperature $T=170$
	and chemical potential $\mu_{B}=0$ \cite{Koch:2005vg}.
	Both HIJING and JETSET results are taken from  \cite{Koch:2005vg}.
    \label{fig::acceptance}}	
\end{figure}
The correlation coefficient $C_{BS}$ is depicted in Fig. \ref{fig::acceptance} as a 
function of the maximum rapidity accepted ($|y|\le y_{max}$).
The analyzed sample consists of central Au+Au events at $\sqrt{s}=200$~AGeV.
For  small acceptance windows around midrapidity, $C_{BS}$ stays roughly constant.
While for a large acceptance window, $C_{BS}$ increases due to the inclusion of 
the fragmentation region with high baryon density.
The different models deviate from each other for large acceptances due 
to differences in the handling of the fragmentation region,  with small rapidity acceptance (relevant for the 
RHIC experiments), HIJING, JETSET and UrQMD yield consistent results.
If the window acceptance covers all produced particles,
$C_{BS}$ has to vanish because of baryon number conservation.

In case a QGP is created, the signal given by the $C_{BS}$ coefficient should survive the hadronic phase.
With a strong enough longitudinal flow,
strangeness and baryon number within a given rapidity range should be frozen in.
The used rapidity window can not be too wide in order to avoid global baryon number and strangeness conservation.
Nevertheless, the acceptance window must be wide enough to avoid smearing due to hadronization.

\begin{figure}
\begin{center}
\includegraphics[width=0.5\textwidth]{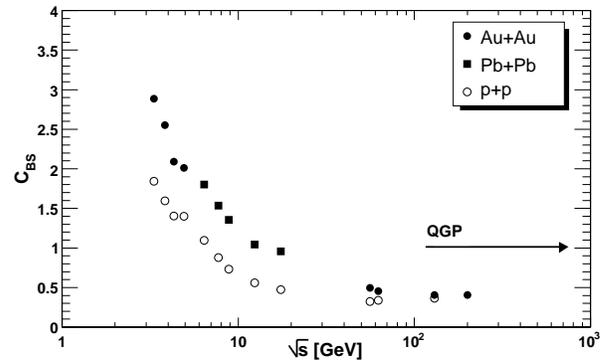}
\end{center}
\caption{Correlation coefficient $C_{BS}$ for central  Au+Au/Pb+Pb (full symbols) and 
minimum bias $p+p$ collisions (open symbols) as a function of $\sqrt{s}$.
The maximum rapidity accepted is $y_{max}=0.5$.
    \label{fig::excitationfunction}}
\end{figure}

Figure \ref{fig::excitationfunction} depicts the energy excitation function of $C_{BS}$ in both p+p 
and centrals Au+Au/Pb+Pb collisions.
As discussed in \cite{Koch:2005vg}, $C_{BS}$ increases with an increase of the baryon chemical 
potential $\mu_{B}$ when going to lower beam energies.
With increasing collision energy, and therefore decreasing $\mu_{B}$, $C_{BS}$ goes down 
to $C_{BS} \approx 0.4$ at the highest RHIC energy available.
Surprisingly, the general trend is the same for both p+p and Au+Au/Pb+Pb.
Measuring the energy dependence of $C_{BS}$ correlation around midrapidity might therefore allow to
map out the onset of the QGP production.

\begin{figure}
\begin{center}
\includegraphics[width=0.5\textwidth]{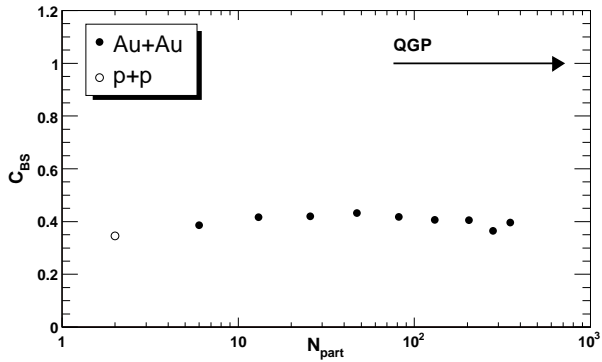}
\end{center}
\caption{Correlation coefficient for Au+Au collisions at $\sqrt{s}=200$ as a function of the number of participants.
    The maximum rapidity accepted is $y_{max}=0.5$.
    Full symbols are the results for Au+Au and the open symbol shows the p+p value.
    \label{fig::participants}}
\end{figure}

The dependence of $C_{BS}$ on the number participants is studied in figure \ref{fig::participants}.
The number of participants is determined via the scaled number of $\pi^{-}$'s in $4\pi$ 
geometry ($N_{part}=0.2 <\pi^{-}>$).
This quantity is proportional to the overlap volume of the colliding nuclei and thus to the number of participants.
The UrQMD model predicts a flat dependence of $C_{BS}$ on centrality.
$C_{BS} \approx 0.4$ from $p+p$ to central $Au+Au$ events.
This is in strong contrast with what is to be expected if the system enters a QGP phase at some centrality.
In this case $C_{BS}$ will increase (whether linearly or as a step function depends on the onset behaviour 
of the QGP phase) 
from peripheral AA or pp towards central AA collisions. This might allow to extract in detail the volume dependence
of the deconfinement transition at RHIC.

To summarize, we have studied the dependence of the baryon-strangeness correlation coefficient 
as a function of the center of mass energy from $E_{lab}=4$~AGeV to $\sqrt s=200$~AGeV for pp and central Au+Au/Pb+Pb
reactions. At $\sqrt s=200$~AGeV we have explored the centrality dependence of the $C_{BS}$ correlation.
$C_{BS}$ is found to decrease from the lower energies towards the top RHIC energy available (here $C_{BS} \approx 0.4$).
For minimum bias $Au+Au$ events at $\sqrt{s}=200$, we predict a flat centrality dependence of $C_{BS}$ near
midrapidity. At the highest RHIC energy the $C_{BS}$ value from the microscopic transport model 
is roughly half the one expected in the case of a QGP.
We suggest to study the energy and centrality dependence of $C_{BS}$ which allow to 
gain information on the onset of the deconfinement transition in temperature and volume.
The CERES/NA49 and STAR experiments should be able to perform these analysis with their accumulated data.

\section*{Acknowledgements}

This work is supported by GSI and BMBF.

\end{document}